# Forecasting Excessive Anesthesia Depth Using EEG α-Spindle Dynamics and Machine Learning

Christophe Sun[*,1], Pierre-Olivier Michel[*,1,2,3], François David[,4], Nathalie Rouach[4], Dan Longrois[5], David Holcman[1,3] *Member, IEEE*

*Abstract- Objectives*: **Accurately predicting transitions to anesthetic drugs overdosage is a critical challenge in general anesthesia as it requires the identification of EEG indicators relevant for anticipating the evolution of the depth of anesthesia.** *Methods*: **In this study, we introduce a real-time, data-driven framework based on α-spindle dynamics extracted from frontal EEG recordings. Using Empirical Mode Decomposition, we segment transient α-spindle events and extract statistical features such as amplitude, duration, frequency, and suppression intervals. We apply these features to train a Light Gradient Boosting Machine (LGBM) classifier on a clinical EEG dataset spanning induction, maintenance, and emergence phases of general anesthesia.** <u>Results</u>: **Our model accurately classifies anesthesia phases with over 80% accuracy and anticipates the onset of isoelectric suppression—a marker of anesthetic drugs overdosage—with 96% accuracy up to 90 seconds in advance.** <u>Conclusion:</u> **The spindle-based metrics provides a non-invasive, interpretable, and predictive approach. This real-time method can be used to forecast unintentional anesthetic drugs overdosage, enabling proactive anesthesia management based solely on EEG signals.** <u>Significance:</u> **This new method is the first to provide a way to prevent too deep anesthesia and its consequence for the well-being of patients after the recovery from anesthesia.**

*Index Terms*—**ElectroEncephaloGram (EEG), α-Spindles, empirical mode decomposition (EMD), Light Gradient Boosting Machine (LGBM), General Anesthesia (GA), Predictive biomarker, Isoelectric Suppression (IES), Depth of Anesthesia (DoA), Cross-Frequency Coupling (CFC), Machine Learning.**

## I. Introduction

Brain rhythms provide an accessible window into neural population dynamics, with transient oscillatory bursts in the α-frequency band (8–12 Hz) frequently observed across various brain states including sleep, general anesthesia (GA), meditation, and sensory stimulation [1], [2], [3]. In particular, *α-spindles* - brief, waxing-and-waning alpha-band oscillations - are a hallmark of stage 2 non-REM sleep [4], [5] and are also commonly seen under GA [6], [7], [8].

While α-spindles are widely reported, their clinical utility and underlying dynamics during GA remain poorly understood. In natural sleep, they are shaped by thalamocortical loops and modulated by both thalamic relay activity and corticothalamic feedback [9], [10], [11], [12]. Spindles are further classified into slow (9–12 Hz) and fast (12–15 Hz) subtypes with region-specific cortical dominance [13], [14], and are signatures of the cortex and thalamus [15].

In contrast, the mechanisms of α-spindle generation under GA and their potential predictive value for monitoring depth of anesthesia (DoA) have yet to be clarified. Of notice, α-spindles often precede pathological EEG patterns such as burst suppression [7], [8], [16], a state associated with anesthetic drugs overdosage that might lead to postoperative cognitive impairment.

In clinical anesthesia, accurate monitoring of brain state is crucial for preventing periods of excessive suppression, particularly burst suppression, a pathological EEG pattern potentially associated with poor post GA cognitive outcomes [17], Standard of care from the societies of anesthesiology recommends avoiding EEG burst suppressions [17], which indicate anesthetic drugs overdosage. Better patient outcomes and safety underlie the need to predict and prevent EEG burst suppressions and not only to correct them once established.

These guidelines reinforce the need for improved real-time EEG biomarkers of impending brain-state transitions.

Current methods for EEG-based DoA monitoring, such as the BIS and PSI indices, rely on aggregate spectral features and do not reliably anticipate transitions to unintentional anesthetic drugs overdosage or isoelectric suppression (IES). Prior signal processing methods -including time-frequency spectrograms [18], wavelet decompositions [19], [20], [21], and multifractal analyses [22], [23], [24]- have not yielded sufficiently predictive markers for early warning of unintentional anesthetic drugs overdosage. These EEG tools lack the resolution to anticipate critical state transitions, such as the onset of isoelectric suppression, limiting translational development.

Here, we hypothesized that α-spindles under GA represent distinct, state-dependent network phenomena that reflect underlying brain dynamics and could serve as predictive biomarkers of anesthetic depth. Here, by correlating spindle duration, amplitude, and frequency across different phases of GA, we aim to generate a statistical machine learning model to analyze an existing human EEG dataset acquired under a comparable anesthetic agent. This approach enables observing distinct EEG signatures associated with α-spindles and their absence during isoelectric burst-suppressions [16], [25]. The impact of this AI algorithm may serve as a predictive biomarker for better anesthesia management.

* contributed equally
1) Group of Data Modeling, Computational Biology and Predictive Medicine, Institut de Biologie (IBENS), Ecole Normale Supérieure, Université PSL, Paris, France.
2) Université Paris Créteil Université Paris-Est Créteil, LAMA, 61 avenue du Général de Gaulle, 94010 Créteil, France.
3) SignalMed+, 229 rue Saint-Honor\'{e}, 75001 Paris, France.
4) Neuroglial Interactions in Cerebral Physiology and Pathologies, Center for Interdisciplinary Research in Biology, Collège de France, CNRS, INSERM, Labex Memolife, Université PSL, Paris, France
5) Département d'Anesthésie, Hôpital Louis Mourier, Assistance Publique - Hôpitaux de Paris, Paris, France
.
(e-mail: david.holcman@ens.fr).

## II. MATERIALS AND METHODS

*Data acquisition and processing statement*

This prospective observational study included 54 young patients (age from 2 to 15) receiving a scheduled elective procedure requiring GA from Louis-Mourier Hospital in Colombes (Agreement NCT06020599 ; APHP230375), France between October 2019 and September 2021, in compliance with the evaluation of clinical practice [26]. The EEG data used were recorded with the Masimo Root® monitor with four frontal electrodes F7, Fp1, Fp2, F8 in the 10-20 system.

We used MATLAB R2021a software to perform the statistical analysis on human patients. Values were expressed in mean ± standard deviation or 95% confidence bounds or median [IQR]. We noted the null hypothesis H0 against the alternative hypothesis H1. The significance level used was $\alpha=0.05$.

*General anesthesia and EEG pre-processing on human*

Patients underwent a standardized general anesthesia procedure, which included three phases (induction, maintenance, and emergence). For induction, the hypnotic agent sevoflurane began to be administered by inhalation with an initial concentration of 6% and a fresh gas flow of 4 l/min under spontaneous ventilation followed by transition to assisted ventilation as soon as possible in order to optimize sevoflurane capture and avoir hypercapnia. Upon an EEG profile of mainly delta waves, the patient had an intravenous line inserted and received opioid analgesics -0.1 $\mu g.kg^{-1}$ sufentanil or 15 $\mu g.kg^{-1}$ alfentanil together with 1 mg/kg of lidocaine. Propofol was injected in one or more bolus doses of 1 mg/kg with 1-minute intervals between two subsequent doses until the appearance of EEG iso-electric suppressions for orotracheal intubation. During the maintenance phase, the hypnotic concentration (sevoflurane) was adjusted depending to obtained continous (qualitatively unfragmented) delta and alpha spectra. Additional drugs (analgesics and muscle relaxants) were given at the discretion of the anesthesiologist as long as the standard of care was respected. Finally, to prepare for the emergence and extubation phase, sevoflurane concentration was typically reduced to values between by 10-20 % during the 5 to 10 minutes preceding complete interruption of sevoflurane inhalation.

For the EEG pre-processing, we first segmented motion-induced mechanical noise seen from the EEG signal within a sliding window of length 10 seconds and 50% overlap. Whenever the signal power within a window $P_i$ exceeded three times the median absolute value

$$MAD = \text{median}(|P_i - \text{median}(P)|), \quad (3)$$

for $i = 1,2,...,N$, where $N$ is the number of windows, we label this event an artifact. The MAD provides a robust estimation of the standard deviation and thus is used to detect outliers in the EEG signal. The artifact segments are then corrected using the Wavelet Quantile Normalization algorithm [27], [28]. This procedure replaces any artifact segment by a physiological surrogate computed from the non-corrupted EEG signal present before and allows to obtain more robust statistics. However, when the artifacts totally overlap with the window of analysis, they are then discarded.

*Spindles segmentation in the EEG*

To define the transient pattern spindle in the EEG from the continuous oscillations, we write the envelope of the signal $x(t)$ such that $x(t) = A(t)\exp(i\phi(t))$, (4) where $A(t)$ accounts for the slow modulation of the envelope compared to the fast oscillating events contained in $\phi(t)$. This envelope can be obtained using the Empirical Mode Decomposition [29], [30]. We thus intuitively define a spindle as the part of the signal between two consecutive amplitude minima that contains at least a global maximum amplitude. We present the steps to segment the spindles:

*Bandpass filter in the frequency of interest* to the signal $x(t)$: we use classical buterworth filtering in α−band frequency range (8-12 Hz).

*Identify local maxima (resp. minima):* this step uses the findpeak routine in MATLAB.

*Remove the effect of sampling noise.* Compute the upper $x_{up}(t)$ (resp. lower $x_{down}(t)$) envelope by using locally weighted linear regression with second-degree polynomial model [31] to smooth the local maxima (resp. minima) points.

*Evaluate the envelope amplitude:*

$$d(t) = |x_{up}(t) - x_{down}(t)|. \quad (5)$$

**Segment two sets of time points** for which the envelope amplitude $d(t)$ have: 1- local maxima above a given threshold $T_{up}$ and 2- local minima below a given threshold $T_{down}$. In principle, we choose $T_{down} = \sigma_x$ and $T_{up} = 3\sigma_x$ where $\sigma_x$ is the standard deviation of the signal $x(t)$

$ST_{up}=\{t_i \mid \exists \varepsilon > 0, \forall t \in [t_i - \varepsilon, t_i + \varepsilon], d(t_i) > d(t) \wedge d(t_i) > T_{up}\}$,

$ST_{down}=\{t_j \mid \exists \varepsilon > 0, \forall t \in [t_j - \varepsilon, t_j + \varepsilon], d(t_j) < d(t) \wedge d(t_j) < T_{down}\}$.

**Define a spindle** as the signal between two consecutive time points in $S_{Tdown}$ such that they contain at least one element in $ST_{up}$:

$SSpindles = [\{x(t) \text{ for } t \in [t_i, t_{i+1}] \mid \exists t_j \in ST_{up}, t_i < t_j < t_{i+1}, \quad (8)$
$(t_i, t_{i+1}) \in ST_{down}\}.$

A single spindle can be further separated into an increasing (waxing) and a decreasing phase (waning): for a spindle labeled $j$, the increasing phase starts from a time $t_j$ where its envelope distance evolves from a minimum to a maximum amplitude $A_j$ for a duration $\Delta_{1,j}$ (Fig. 1D). The decreasing phase starts from the spindle maximum amplitude until the next minimum $t_{j+1}$, for a duration $\Delta_{2,j}$. The spindle total duration associated with label $j$ is thus $\Delta_j = \Delta_{j,1} + \Delta_{j,2}$.

Note that the spindle duration is the sum of the two exponential distributions (increasing and decreasing phases). Since the exponential decaying factors are not very different, the total distribution is obtained by a convolution of the two exponentials with the same factor leading to an Erlang-2 distribution, showed in Fig. 1.

We used fragmentation as a metric to estimate the proportion of time spent without significant spindles in a window. We call such occurrence an off area. Nonetheless, if the time duration of an off area is lower than a threshold, this signal is not counted. This means that only long distance between significant spindles will be taken as off time. Computing this metric on sliding windows over time allows to track if the time spent with no spindles is evolving. Briefly to compute the fragmentation, we take all segments bellow 0.2 of the 80$^{th}$ percentile of the smooth power time series, we use joint segment separated by less than 0.15 s and keep only segment longer than 0.4 s. The fragmentation value is the sum of the segments' lengths over the window length. We computed the block entropy of the band-pass signal with a 2-grams. Meaning the alpha wave signal is divided into blocks of non-overlapping sequence of 2 consecutive points. Then the entropy of the distribution of those blocks is computed.

*Spindle types classification*

To automatically classify spindles within three categories that we define as dominant slow, dominant fast, and mixed frequencies, we calculate the instantaneous frequency in the spindles region over time. We then label the spindle as slow if the proportion of the time spent below 10 Hz against the total duration $p_{f<10}$ is greater than 60%. On the other hand, when the proportion of the time spent above against the total duration $p_{f>10}$ is greater than 60%, we label the spindle as fast. Finally, when both proportions are within 40% and 60%, we label the spindle as mixed.

*Fragmentation to evaluate significant absence of spindles*

We used fragmentation as a metric to estimate the proportion of time spent without significant spindles in a window. We call such occurrence an off area. Nonetheless, if the time duration of an off area is lower than a threshold, this signal is not counted. This means that only long distance between significant spindles will be taken as off time. Computing this metric on sliding windows over time allows to track if the time spent with no spindles is evolving. Briefly to compute the fragmentation, we take all segments bellow 0.2 of the 80$^{th}$ percentile of the smooth power time series, we use joint segment separated by less than 0.15 s and keep only segment longer than 0.4 s. The fragmentation value is the sum of the segments' lengths over the window length.

*Block Entropy*

We computed the block entropy of the band-pass signal with a 2-grams (Shannon, 1948). Meaning the alpha wave signal is divided into blocks of non-overlapping sequence of 2 consecutive points. Then the entropy of the distribution of those blocks is computed.

*Power*

Given a signal s(t) of length N, we define its power as:

$P = \text{median}([s(t_i)^2]_{i\in[1,N]})$

The use of the median instead of the common definition using the mean allows to be more robust to possible outliers

*Main frequency*

Given a signal in a frequency range $[f_0, f_1]$, we define as main frequency the frequency corresponding to the center of mass of the signal Power Spectral Density.

*GA phase prediction*

For predicting the phases of GA, 5 EEG samples of each 52 patients of 1-minute length, and for each of the three GA phases - induction, maintenance and emergence were used as dataset. The α-spindles' characteristics were extracted: the mean α-spindles duration, amplitude, frequency and number, the relative proportions of slow, mixed, and fast α-spindles and the duration and number of α-suppressions. For each GA phase and the IES to predict, we trained various models, among which the Light Gradient Boost Model reported here provided fast and accurate results. The dataset was split into subdataset samples 2/3-1/3 proportion trials for training and testing. A cross-validation using 4 data shuffling was used and hyperparameters of each model scanned for optimization (using Scikit-learn Python library). The classifier provides prediction 0 or 1. True Positive and True Negative proportions are reported in confusion matrices for each GA phase. To explain the model performance, we used the Shap explainer library [32], which allowed us to rank each EEG characteristic by order of importance of the averaged values and to attribute an importance SHAP value (Shapley Additive) of that characteristic to each EEG sample (Fig. 4A).

*First IES occurrence prediction*

To predict the occurrence of a first IES, 59 samples of 4 min EEG data were labeled. They are marked as 0 if 90 s just after the sample there is no EEG, and classified as 1 if there is an IES. Each 4-min sample is divided into 90 s sliding windows with a 5 s step for a total of 30 subdivisions. On each sliding window the α, δ powers and main frequencies are computed, as well as α-related metrics - block entropy, number of spindles, proportion of low, mixed and high spindles of the α wave are added. The variables are then smoothed to get rid of small fluctuations on the 30 points using a first order Savistky-Golay filter. This smoothing method was selected for its edge preservation properties. Finally, a linear fit is made to extract for each variable the value at origin and trend. A sample is consequently converted from a raw EEG recording of 4 min to a 1D array of values at origins and trends suitable for a classifier training. An LGBM classifier is trained on the dataset of 59 samples with 39 label 0 and 20 label 1. We used a stratified train-test splitting to compensate the imbalance of the dataset. A 5-fold cross-validation is implemented to help prevent overfitting and tune hyperparameters. The same SHAP routine is done to find the importance degree of the different features.

III. RESULTS

**Datasets for Thalamocortical Analysis**

Two human cohorts (pediatric and adult) were under sevoflurane and propofol per standard of care [26]. A separate human cohort was used for sleep data acquired from an online database [33].

**Diversity of α-spindles in patient EEG during clinical general anesthesia**

*Alpha (α)-spindles* are the underlying patterns generating the α band on the spectrograms of several DoA monitors. This pattern could either be endogenous or reflect a message to be decoded. It could encode early signs of cortical-thalamic decoupling in human patients recorded non-invasively under clinical anesthesia. To explore this, we analyzed α-band EEG oscillations from patients under GA and focused on the temporal architecture of α-spindles as potential biomarkers of brain state and circuit engagement. The EEG during stable GA exhibits prominent α-band activity consisting of discrete spindle-like bursts followed by periods of quiescence or succession of α-spindles (Fig.1A,B). These episodes mirror the canonical waxing and waning envelope structure of spindles, reminiscent of those in natural sleep and rodent recordings. The spectrogram (Fig.1C) reveals a continuous alternation between high- and low-intensity epochs at different frequencies, suggesting non-stationary oscillatory dynamics across frequencies. To isolate these events, dealing with the non-stationarity of the EEG signal [27], [34], we develop a segmentation of α-spindles using the Empirical Mode Decomposition (EMD) approach [29] to compute the upper and lower envelopes (Fig.1D). This allowed reproducible segmentation of waxing and waning phases based on local envelope curvature.

The power spectrum averaged across the patient cohort revealed a robust peak near 10 Hz (Fig.1E), anchoring these α-spindles firmly within the classical α rhythm range. Yet, the dynamic structure within individual spindles is anything but uniform. The distributions of waxing and waning durations were well-fit by single exponential functions, suggesting a stochastic process modulated by intrinsic circuit relaxation time constants. These results demonstrate that even in human EEG, α-band activity under GA is composed of discrete, non-periodic events with internally structured dynamics. The variability in α-spindle morphology may reflect the balance of cortical and thalamic contributions-with shorter, sharper spindles reflecting stronger cortical amplification and longer, flatter ones pointing to more balanced circuit dynamics.

We found that the distributions of waxing and waning phases durations can be both parameterized by a single exponential (Fig.1F, 1$^{st}$ row here for a single patient) $f_{\Delta 1}(t) = \lambda_1 \exp(-\lambda_1 t)$, and $f_{\Delta 2}(t) = \lambda_2 \exp(-\lambda_2 t)$ respectively, where fits yielded to $\hat{\lambda}_1 = 1.2 \pm 0.1 s^{-1}$ ($R^2 = 0.96$) and $\hat{\lambda}_2 = 1.6 \pm 0.3 s^{-1}$ ($R^2 = 0.93$). Finally, the total duration Δ is fitted by the convolution

$$f_\Delta(t) = \lambda^2 t \exp(-\lambda t) \quad (1)$$

yielding to $\hat{\lambda} = 1.3 \pm 0.3$ ($R^2 = 0.92$) (Fig.1F, 2$^{nd}$ row). This fits the hypothesis that the increasing and decreasing phases duration are statistically independent. Furthermore, we fitted the distribution of the inter-spindles duration $\Delta_{sup}$ with a single exponential curve, leading to $\hat{\lambda}_2 = 0.5 \pm 0.2 s^{-1}$ with $R^2 = 0.95$ (Fig.1F, 2$^{nd}$ row). Analysis of the patient cohort (n=54), showed the distributions of the estimated parameters across individuals (Fig.1G, left). The exponential decay factors estimated on the different durations ($\lambda_1 = 0.93$ [0.89-0.99 IQR (Inter Quartile Range)] $s^{-1}$, $\lambda_2 = 0.94$ [0.89-1.01] $s^{-1}$, $\lambda = 1.07$ [1.01-1.13] $s^{-1}$ and $\lambda_{sup} = 0.57$ [0.40-0.87] $s^{-1}$), indicate that the waxing and waning phases follow similar structure-functions in the patient cohort. Lastly, we collected the maximum amplitude A of the spindles and the distance $|x_{up} - x_{down}|$ between the upper $x_{up}$ and lower $x_{down}$ envelopes. Both the amplitude and the distance $|x_{up} - x_{down}|$ are fitted with a Γ−distribution

$$G_{a,b}(x) = \frac{x^{a-1} \exp(-\frac{x}{b})}{\Gamma(a) b^a}, \quad (2)$$

where $\Gamma(x) = \int_0^\infty t^{x-1} e^{-t} dt$ is the Γ−function. The parameter *a* represents the shape of the distribution while *b* is for its width. The optimal parameters for a single patient were $\hat{a}_{amp} = 1.4 \pm 0.1$, $\hat{a}_{env} = 1.8 \pm 0.1$ in arbitrary units and $\hat{b}_{amp} = 5.1 \pm 0.7 \mu V^{-1}$, $\hat{b}_{env} = 4.4 \pm 0.1 \mu V^{-1}$ (Fig.1F, 3$^{rd}$ row). In contrast, the Γ−distribution on the maximum amplitude and envelope average amplitude had large variability across the cohort (Fig.1G right): $a_{amp} = 2.26$ [1.96-2.55] a.u., $b_{amp} = 4.39$ [3.25-5.89] $\mu V^{-1}$ (resp. $a_{env} = 2.20$ [1.78-2.60] a.u., $b_{env} = 4.70$ [3.50-6.77] $\mu V^{-1}$).

We applied the present segmentation to α-spindles recorded during sleep in Fig. SI1, revealing statistical similarities that could thus reflect a potential use of those spindles as predominantly associated with sleep stage 2, which correspond to an intermediate sleep depth.

**Induction, maintenance and emergence phases are characterized by distinct α-spindle dynamics**

Brain rhythms like α-spindles express stochastic variability in shifting circuit configurations within and across patients. However, we speculated that these parameters could serve as indicators of clinical phases and help in decision-making in controlling GA. Under GA, we examined transitions between GA phases to study how population synchrony adapts to externally controlled perturbations. In the same patient cohort, we pooled α-spindles dynamics (Fig.2A-C) and δ-waves (Fig. SI2) for each of the three main stages of anesthesia: 1) induction just following an intravenous propofol injection (bolus 2 mg.kg$^{-1}$), 2) the maintenance with inhaled sevoflurane (between 1% and 5%) and 3) emergence once GA is stopped. α-suppressions were significantly shorter during maintenance and longer during induction (9.0, 1.3, 5.1 s for induction, maintenance and emergence, respectively; p<0.05, Wilcoxon paired test, Fig.2B-C top rows).

# Human α-Spindles EEG Characteristics During Clinical Anesthesia

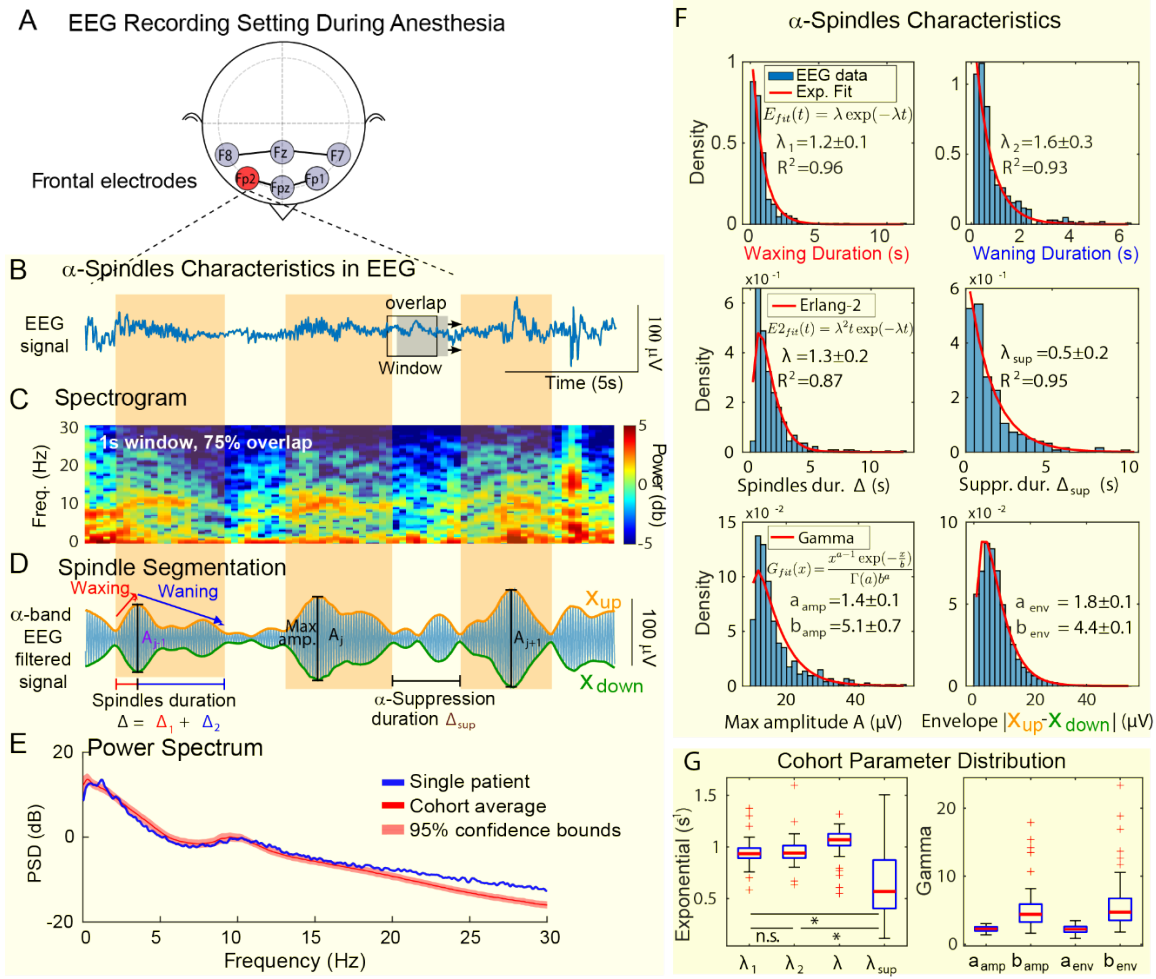

**Fig. 1 Spindles dynamics and statistics obtained in human EEG recordings during clinical general anesthesia (A)** Electrodes positioning on the scalp. **(B)** Unfiltered EEG signal showing transient spindles (yellow rectangles). **(C)** Time-frequency representation (spectrogram) using 1 second time window size and 75% overlap. **(D)** Spindles (yellow rectangles) present in the filtered EEG in the α-band range and their decomposition based on the maxima and minima of the distance between the upper (yellow) and lower (green) envelopes using the EMD technique (see Methods). Spindles are separated into an increasing (blue), called waxing and decreasing (red) waning phases by a maximum amplitude (violet). The duration between two α-spindles marks a period of α-suppression (white space). **(E)** Power spectrum densities on a single patient (blue) and on a patient cohort (n=54) average with its 95% confidence bounds (red and shaded). **(F)** Single patient α-band spindles statistical distributions. First row: Probability density function (pdf) of spindles increasing phase duration $\Delta_1$, decreasing phase duration $\Delta_2$, both fitted with an exponential curve (red). Second row: Pdf of the total duration fitted with an Erlang-2 distribution, which is a convolution of two exponential (red) and the α-spindles suppressions duration $\Delta_{sup}$. Third row: Pdf of the spindle maximum amplitude $A_k$ and the envelopes distance $|x_{up} - x_{down}|$, both fitted with a Gamma distribution (red). Estimated values are expressed with 95% confidence bounds. **(G)** Boxplots of the estimated parameters over the patient cohort (n=54). Left: the exponential decay factors estimated on the different durations $\lambda_1$ = 0.93 [0.89-0.99 IQR] (increase), $\lambda_2$ = 0.94 [0.89-1.01] (decrease), $\lambda$ = 1.07 [1.01-1.13] (total duration) and $\lambda_{sup}$ = 0.57 [0.40-0.87] (suppression) in $s^{-1}$. Right: the gamma shape (in arbitrary units) and scale (in $\mu V^{-1}$) on the maximum amplitude ($a_{amp}$=2.26 [1.96-2.55], $b_{amp}$=4.39 [3.25-5.89]) and the envelopes distance ($a_{env}$=2.20 [1.78-2.60], $b_{env}$=4.70 [3.50-6.77]). * indicate significant difference (p<0.001) between the parameters $\lambda_1$, $\lambda_2$ vs $\lambda_{sup}$. The difference was not significant for $\lambda_1$ vs $\lambda_2$ (p=0.25).

Maintenance was marked by more frequent but briefer α-suppressions and a higher number of α-spindles (Fig.2C 2nd and 3rd rows), reflecting a higher fragmentation of those episodes and persistent engagement of the thalamocortical system. Features of α-spindle such as duration, amplitude, and frequency were significantly different between most GA phase pairs (Fig.2B,C last 3 rows). The largest differences were smaller α-spindle durations during induction, some higher α-amplitudes during maintenance, and higher frequencies during emergence. Most strikingly, the α-spindles' numbers, durations, and amplitudes showed a transient decrease and then increase after the propofol bolus (Fig.2C 3rd to 5th rows), while being steady during maintenance, and showing a steady decrease during emergence. The average α-spindle frequency was strikingly steady within each phase of anesthesia and significantly differed among each GA phase (Fig.2B, C, last row), reaching a maximum during emergence. In contrast to α-spindles, we observed a transient increase, then decrease in the mean duration and amplitude of δ-wave during induction, stability during maintenance (Fig. SI2A-B).

## Human α-Spindles Characteristics across Phases Anesthesia

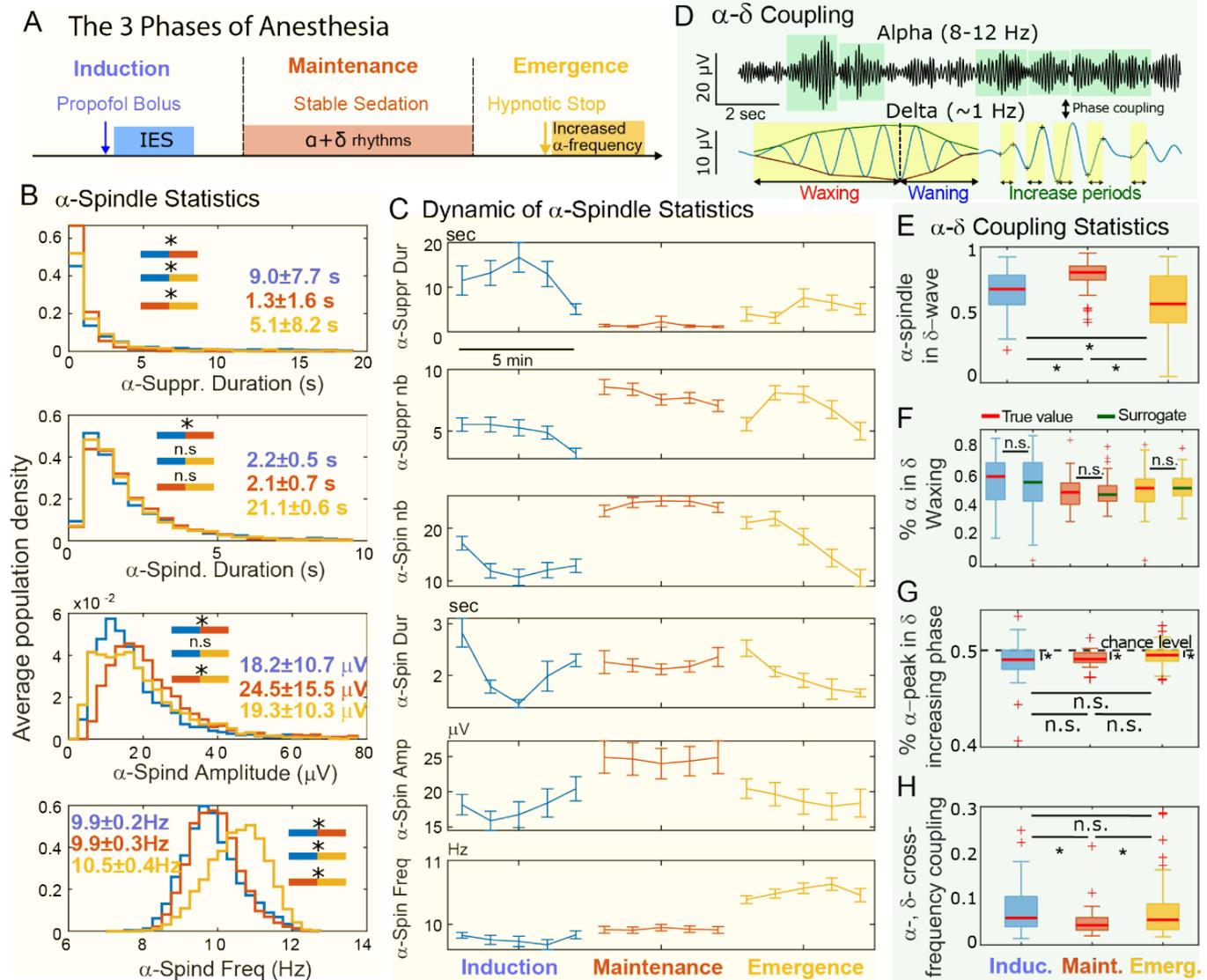

Fig. 2. *α*-**spindle statistics for the different phases of anesthesia.** (A) Timeline of the different anesthesia phases: P1) after a bolus of propofol during induction (blue), P2) during the maintenance phase (red) and P3) after the hypnotic injection is turned off in the emergence phase (yellow). (B) Average population probability density functions of the α-suppression and α-spindle parameters (duration, amplitude and mean frequency) for the different phases of anesthesia described in (A). Values shown on the figure correspond to the population mean ± standard deviation. * indicate significant difference between the population averages across anesthesia phases. (C) Time evolution of α-suppression and α-spindle parameters for each anesthesia phase described in (A). The population average and standard error were calculated at 5 intervals of 1 minute each. (D) Spindles extracted from an example of EEG filtered within both the α (green rectangle) and δ (yellow) frequency bands. (E) Mean proportion distributions of α-spindles within δ−waves on human patients (n=54). * indicate significant difference between the three anesthesia phases (p<0.05). (F) Mean proportion distributions of α-spindles in δ−wave waxing envelope. True data were shown on the left side of each phase (median in red). Surrogate data (uniform sampling for each α-spindle within the corresponding δ−wave interval) on the right side of each phase (median in green). The differences were not significant between the true data and the surrogate for all phases. (G) Mean proportion distributions of α-spindles on δ−waves increase periods. The differences were significant for all phases against the 0.5 chance level (p<0.05). (H) Distributions of coupling index from α and δ frequency bands coupling. Difference was significant between maintenance and the other two phases (p<0.05) but not for induction vs emergence (p=0.44)(n=54 in all statistical tests).

The emergence phase was characterized by an overall decrease of those parameters. In this cohort, the number of IES was small and only present during induction (Fig. SI2C). Like with α-spindles, δ-waves, and IES show significantly different distribution of their characteristic values between GA phases. As part of a single network dynamics, the α-spindles and δ-waves co-occur and arise from over-lapping yet distinct networks. We examined their interactions across GA phases (Fig.2D). The probability of α-spindles within δ-waves was significantly higher during maintenance 0.82 [0.77-0.87 IQR] compared to induction 0.69 [0.57-0.80] and emergence 0.57 [0.43-0.79] (Fig.2E; p<0.05, Wilcoxon paired test). Yet the phase relationship was stochastic, showing no consistent preference for δ-waves waxing phases 0.58 [0.42-0.67] during induction, 0.47 [0.39-0.54] during maintenance, and 0.5 [0.41-0.56] during emergence (Fig.2F) compared to temporally shifted surrogate data. Still, we found a modest but significant under-representation of spindles during δ-wave increasing periods compared to the decreases across all phases (Fig.2G, p<0.05, Wilcoxon test) for all phase (P1: 0.49 [0.480.50], P2: 0.49 [0.48-0.50] and P3: 0.50 [0.49-0.50]). These results suggest a subtle alignment of excitability windows with nested rhythms. However, there was no significant difference between GA phases.

Finally, we probed cross-frequency coupling (CFC) between α and δ bands, a hallmark of healthy and pathological brain dynamics [35]. We tested whether the GA phases impact the frequency coupling between. Surprisingly, the coupling index [36], [37] was weakest during the maintenance phase (0.03 [0.02 to 0.05 IQR]) (Fig.2H), despite a maximal α-δ co-occurrence. This paradox may reflect a functional decoupling between nested rhythms under deep sedation, where co-expression does not require or result in phase alignment. To conclude, the maintenance phase is characterized by a maximal co-occurrence of α-spindle and δ−wave (Fig.2E) and a minimal CFC (Fig.2H) between α and δ frequency bands. Together, these results suggest that α-spindles encode more than transient oscillatory events. We posit they track state transitions, reflect changing circuit excitability, and interact with slower rhythms in a context-dependent manner.

### α-Spindles categories sequence of appearance reveal a random process

Oscillatory patterns in the EEG emerge not only from anatomical structures but also from dynamic transitions between functional states. Given the striking recurrence and phase-specific modulation of α-spindles, we asked whether their intrinsic features and their moment-to-moment fluctuations could serve as indicators of the brain's state along the anesthesia trajectory. Although the mean α-spindle frequency overlapped between induction and maintenance, this average masked importance of heterogeneity. We performed a finer-grained spectral analysis. Using a spectral decomposition analysis, we categorized spindles into three types: 1-fast dominant frequency, 2-slow dominant and 3-equally mixed frequencies (Fig.3A and Methods). These results suggest that α-spindles may reflect coexisting oscillatory modes rather than a unimodal rhythm.

Then we tracked the relative abundance of each spindle type across the 3 GA phases (Fig.3C). Slow-dominant spindles prevailed during induction and maintenance, while emergence was marked by a relative increase in fast-dominant spindles. This phase-specific shift in spindles subtypes supports the idea that α-spindles are a heterogeneous ensemble of oscillatory motifs whose distribution changes with the underlying network excitability and thalamocortical synchrony. This also suggests that the classification could further help differentiate GA phases. Lastly, we wondered whether the sequence of slow, fast, and mixed spindles was deterministic of a specific pattern. However, we found non-significant p-values for respectively 90%, 94%, and 84% of the population (Fig.3C), as revealed by the Wald-Wolfowitz runs tests [38] across our patients cohort.

This suggests that the three types of α-spindles are generated in a random order.

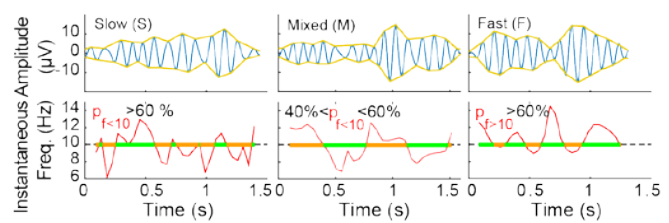
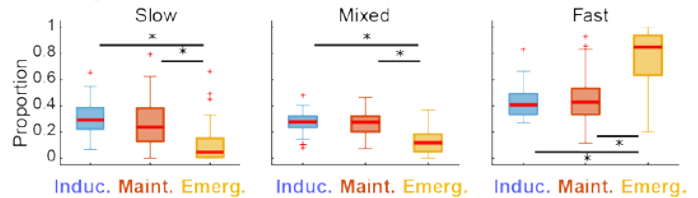
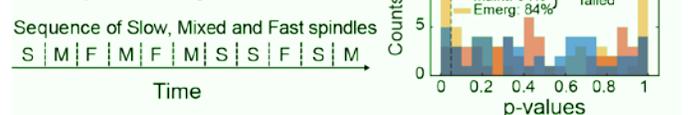

Fig 3: **Spindles classification, encoding and comparison across GA. (A)** 3 examples of α-spindles (1st row) and their instantaneous frequency (2nd row). Slow α-spindles (left) are defined by a percentage (>60%) for duration for which the instantaneous frequency is below (yellow) 10 Hz and the total spindle duration. Fast α-spindles (right) are defined when instantaneous frequencies above 10Hz cover more than 60% of total spindle duration. The rest of α-spindles are considered mixed. **(B)** Proportions of slow (0.29, 0.24, 0.04), mixed (0.24, 0.24, 0.12) and fast (0.41, 0.43, 0.85) spindles in the Induction, Maintenance and Emergence GA phases. • indicate significant differences for Induction and Maintenance vs. Emergence in all spindles types (p<0.001) and for Induction vs. Maintenance for slow spindle type proportions. **(C)** Left: Example of a sequence of 11 spindles. Right: Histogram of the p-values for the deterministic sequence hypothesis (Wald-Wolfowitz test, n=54 patient) for each anesthesia phase. 5% significance level was used to calculate the percentage of patients for which the hypothesis is rejected.

### Inferring GA phase from α-spindle ensemble statistics

From a translational perspective, a major clinical goal is to predict the undergoing phase of GA from limited EEG data to ensure the expected/intentional DoA. We trained machine-learning classifiers using 1-minute EEG samples to predict GA phases based on α- and δ-band features (Fig. 4A-C). The best-performing model was a Light Gradient Boosting Machine (Light GBM). Using twelve previously studied features, the model reached 75% accuracy for the induction phase, 83% for the maintenance, 78% for emergence phases, respectively (Full confusion matrices in Fig. 4B). Notably, specificity was also high across phases, suggesting the model was not biased by class prevalence. The correct prediction of being outside the phase was equally good (88% for not being in Induction, 79% for not being in maintenance, 86% for not being in emergence). To probe the relative contribution of rhythm types, we removed α-spindle indicators and re-trained the model using only δ-wave-related features. Performance decreased across all phases – most strikingly for induction (75% to 67%) and maintenance (83% to 71%, Fig. 4B). This result highlights the determinant role of α-spindle characteristics in tracking GA state. Conversely, removing δ-band features decreased classifier performance for induction (75% to 62%) and emergence (78% to 72%) but not maintenance, with a true positive prediction which stayed at 83%. The percentage of correct prediction of not being in the phase is even more robust to the sole presence of α-spindle features compared to δ-band features. These results suggest that α-spindles carry the most discriminative information during this clinically critical phase. In addition, we ranked the variables for each GA phase that influenced the classifier's decision-making on classification (SHAP values in Fig. 4 C). During maintenance, the duration of α-suppressions, the proportion of fast α-spindles and spindle count were the top predictors. During emergence, spindle frequency, amplitude, and fast-spindle proportion consistently emerged as a key feature. Hence, the proportion of fast- α-spindles appears important in the correct identification of the 3 phases of GA.

**Predicting first IES occurrence**
Finally, we tested whether the α-spindle-derived parameters could predict transitions into deep sedation. Using the adult clinical cohort, including elderly subjects, we trained an LGBM classifier to predict suppressions 1 min and 30 sec before their occurrence (Fig. 4 D-F). The model used a best fit on the following EEG signal features: α power, block entropy, α fragmentation index, spindle count within a sliding window, fraction of fast and slow spindles, mean frequency of the α-band, power in the δ-band. Training was done on EEG segments labeled 0 (no upcoming suppression) and 1 (suppressions occurring 90 seconds later) (Fig. 4 E). Interestingly, the model achieved 96% accuracy with a 100% true positive rate and a 92% true negative rate (see confusion matrix in Fig. 4F). When the same classifier was trained without α-related features, performance dropped dramatically to 33% true positives and 67% true negatives. The SHAP analysis revealed that the 6$^{th}$ most important variables are: the fragmentation, the α power, trend of α power, trend of fragmentation, block entropy and delta main frequency. This indicates the importance of tracking the α-spindles dynamics to prevent IES as the 5 first variables are directly linked to it. These data highlight the critical role of α-spindle statistics in anticipating profound changes in brain state. These findings suggest that the early dynamics of α-spindles offer advanced warning of an impending transition into deep sedation, raising the possibility that preemptive dose adjustment to avoid IES.

## IV. DISCUSSION

Our approach enabled observing distinct EEG signatures associated with α-spindles and their absence during isoelectric burst-suppressions. We discovered distinct properties of α-spindle dynamics related to DoA. We leveraged in vivo α-spindles signatures to develop an AI algorithm that can predict GA phases and the appearance of isoelectric suppressions.

Our predictive biomarker can predict anesthesia phase transitions with up to 83% success and reliably anticipates the onset of isoelectric suppressions up to 90 seconds in advance. This predictive capability may open the door to proactive anesthesia management (i.e., avoiding and not only correcting anesthetic drugs overdosage) by preemptively adjusting dosages to avoid unintentional anesthetic drugs overdosage and mitigate associated risks. Overall, our study characterizes α-spindles in direct relation to their dynamics across GA phases, providing a novel approach to predict and adjust anesthetic drugs dosage for improved patient care. In patients under GA, using this time-resolved statistical structure, we trained a machine learning model to forecast transitions toward deep anesthesia, specifically the onset of IES, up to 90 seconds in advance using EEG data alone on patient cohorts.

**Statistics of α-Spindles Across Anesthesia Phases**
Our analysis of α-spindles features across the three phases of GA reveals distinct statistical signatures that reflect evolving neural dynamics. As shown in Figure 2C, spindles numbers were lowest during induction, increased during maintenance, and decreased again during emergence, while spindles amplitudes peaked also during maintenance. The mean α-frequency remained stable within each phase but shifted across phases, increasing most notably during emergence but also during maintenance. Suppression intervals between spindles (α-suppressions) were shortest and most frequent during maintenance, suggesting a fragmented but persistently active thalamocortical loop. These observations are consistent with prior reports showing that α oscillations under propofol anesthesia are phase-specific and modulated by cortical-thalamic interactions [7], [16]. However, while earlier studies primarily characterized spectral power changes, our study provides a time-resolved statistical analysis of spindles morphology and recurrence, linking them directly to transitions in brain state. The use of exponential and Erlang distributions to model spindle durations supports the view that α-spindles emerge from stochastic processes with memoryless waxing and waning components. These features offer a robust basis for real-time classification of anesthesia depth and forecasting of critical transitions, such as the onset of isoelectric suppression.

**Predicting unintentional anesthetic drugs overdosage minutes in advance**
The gold standard for brain monitoring during GA relies on commercial devices such as the Bispectral Index (BIS, Medtronic) or the Patient State Index (PSI, Masimo). This monitoring offers a snapshot of brain activity by extracting composite features for the alpha and delta bands. These indices are effective for assessing the present brain state, but they fall short in forecasting rapid transitions to unintentional anesthetic drugs overdosage, a state associated with potential synaptic injury and cognitive dysfunction post-surgery [39]. Critically, preventing over-dosage requires anticipation, not just observation followed by correction. This remains an unmet challenge in clinical anesthesia.

Here, we demonstrate that this prediction is achievable by harnessing spindle statistics as shown in Fig 4. Now we can predict the transition toward the anesthetic drugs overdosage (deepest anesthesia phase) with a lead time of approximately 90 seconds. The clinical implications are substantial. Prolonged "deep anesthesia" can alter synaptic structure and reduce dendritic spine density in animal models [40], which may underlie the post-anesthesia cognitive dysfunction in humans. The ability to preemptively detect drift toward this state enables a paradigm shift from reactive (correction of anesthetic drugs overdosage resulting in IES) to proactive anesthesia (prevention of anesthetic drugs overdosage and of IES) monitoring. The present approach paves the way for online signal processing, spindle segmentation, and statistics that could be used to pilot anesthesia in real-time based on the brain activity measured by EEG only.

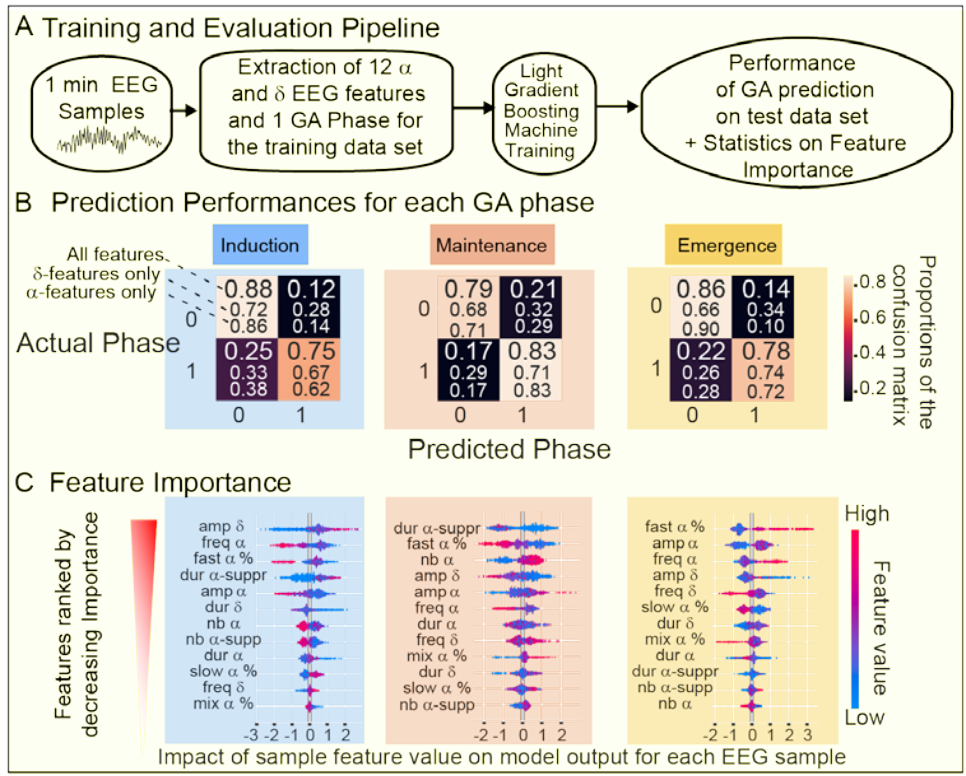

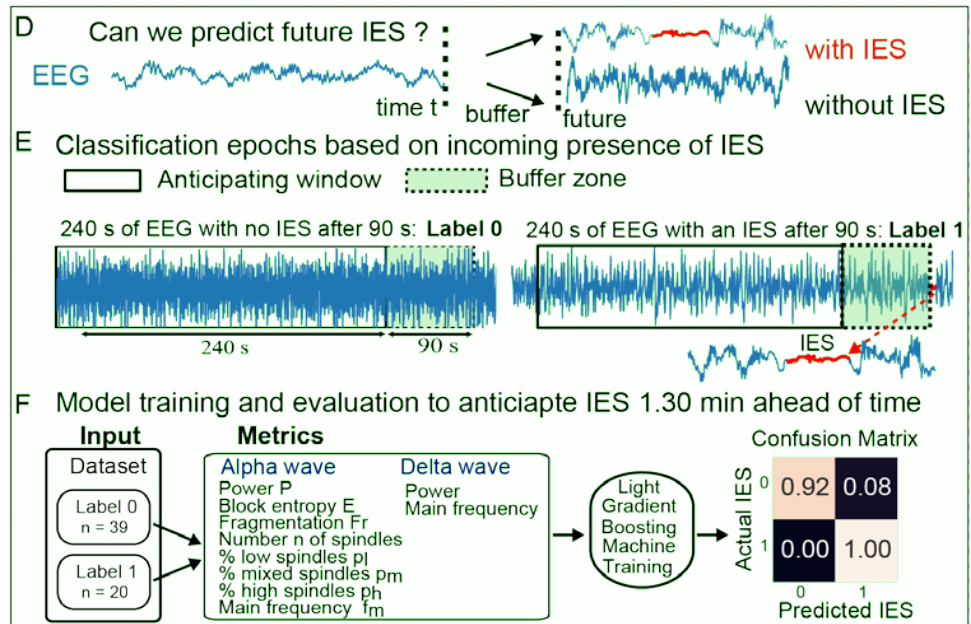

Fig 4. *α*-spindles based prediction of GA phases and predictive anticipation of IES. A) Analysis pipeline for the LGB model of GA phase prediction including training, evaluation and feature importance extraction from 1-min EEG sample. B) Confusion matrices of predicted GA phase from EEG samples (n=270). Full scale model including all set of variables - large font, 1st row. Excluding the α-related features (small font, 2nd row) reduces performance. Excluding the δ-related features (small font 3rd row in each case) tends to reduce performance, less than the exclusion of the α-related features in most of cases. C) Relative contribution of all variables to the prediction. The importance of each feature (color coded) and the direction of their contribution for each sample for this variable is represented for each sample (dot) as the SHAP value (x-axis). D) Schematic question for the anticipation of the presence or absence of IES ahead of time. E) Dataset construction and label setting. A sample is labelled 0 if there is one IES at the end of a 90-second sample following a 4-min sample. It is labelled 1 if there is an IES. As an anticipation time buffer, the 90 second sample is excluded from feature extraction. Feature values are extracted only for the 4-min sample preceding the 90-second interval and the model is trained on the features values of those 4-min EEG samples and the labels. F) The extraction of indicators include average values and their temporal evolution based on a linear fit of the smoothed variable resulting in the value at the origin and the slope of the time varying variable. The confusion matrix indicates a proportion of 1 of truly anticipated IES and 0.92 of truly absent IES.

The next translational step is clear: develop lightweight, real-time algorithms that continuously segment EEG, extract spindle parameters, and compute predictive probabilities of IES onset with sufficient time for the practitioner to prevent deep sedation. Such a system could be implemented using the existing monitoring EEG electrodes. Importantly, this would transform EEG from a passive monitoring tool into an active controller of DoA, adjusting dosage dynamically in response to imminent shifts detected at the network level. By decoding the language of spindles, we move closer to precision EEG monitoring, where cortical and thalamic signatures guide clinical decision-making in real-time and ahead of time. Such a step will also allow us to train the present generation of deep-learning models to decode the brain EEG activity during anesthesia.

*Ethics statement*
Ethical review and approval was not required for the study on human participants in accordance with the local legislation and institutional requirements. Written informed consent to participate in this study was provided by the participants and/or the participants' legal guardian/next of kin. The EEG data used in the study were collected from Louis-Mourier Hospital in Colombes, France between October 2019 and September 2021, in compliance with the evaluation of clinical practice and included a total of 77 patients (50 children and 27 adults), along with their demographic factors and other clinical information. Patients with incomplete or corrupted data were not included in the study.

**Acknowledgments**
D.H. research is supported by ANR grants AstroXcite and AnalysisSpectralEEG and the European Research Council (ERC) under the European Union's Horizon 2020 research and innovation programme (grant agreement No 882673). N.R. research was supported by ANR (AstroXcite) and Labex Memolife.

**Competing interests**
The Authors declare no competing financial interests.

**Data availability**
The datasets generated and analysis in the current study are available from the corresponding authors upon reasonable request.

REFERENCES

[1] G. Buzsaki, *Rhythms of the Brain*, 1ʳᵉ éd. Oxford ; New York: OUP USA, 2011.
[2] F. Travis, « Autonomic and EEG patterns distinguish transcending from other experiences during Transcendental Meditation practice », *International Journal of psychophysiology*, vol. 42, nº 1, p. 1-9, 2001.
[3] B. R. Cahn, A. Delorme, et J. Polich, « Event-related delta, theta, alpha and gamma correlates to auditory oddball processing during Vipassana meditation », *Social cognitive and affective neuroscience*, vol. 8, nº 1, p. 100-111, 2013.
[4] A. L. Loomis, E. N. Harvey, et G. Hobart, « Potential Rhythms of the Cerebral Cortex During Sleep », *Science*, vol. 81, nº 2111, p. 597-598, juin 1935, doi: 10.1126/science.81.2111.597.
[5] A. Kales et A. Rechtschaffen, « A manual of standardized terminology, techniques and scoring system for sleep stage of human subject ». Neurological Information Network, 1968. Consulté le: 30 novembre 2024. [En ligne]. Disponible à: https://cir.nii.ac.jp/crid/1370004237604151300
[6] V. A. Feshchenko, R. A. Veselis, et R. A. Reinsel, « Propofol-Induced Alpha Rhythm », *Neuropsychobiology*, vol. 50, nº 3, p. 257-266, sept. 2004, doi: 10.1159/000079981.
[7] P. L. Purdon *et al.*, « Electroencephalogram signatures of loss and recovery of consciousness from propofol », *Proceedings of the National Academy of Sciences*, vol. 110, nº 12, p. E1142-E1151, 2013.
[8] P. L. Purdon, A. Sampson, K. J. Pavone, et E. N. Brown, « Clinical electroencephalography for anesthesiologistspart I: background and basic signatures », *Anesthesiology: The Journal of the American Society of Anesthesiologists*, vol. 123, nº 4, p. 937-960, 2015.
[9] R. Morison et D. Bassett, « Electrical activity of the thalamus and basal ganglia in decorticate cats », *Journal of Neurophysiology*, vol. 8, nº 5, p. 309-314, 1945.
[10] D. Contreras, A. Destexhe, T. J. Sejnowski, et M. Steriade, « Spatiotemporal patterns of spindle oscillations in cortex and thalamus », *Journal of Neuroscience*, vol. 17, nº 3, p. 1179-1196, 1997.
[11] Y. B. Saalmann, M. A. Pinsk, L. Wang, X. Li, et S. Kastner, « The pulvinar regulates information transmission between cortical areas based on attention demands », *science*, vol. 337, nº 6095, p. 753-756, 2012.
[12] J. G. Klinzing, N. Niethard, et J. Born, « Mechanisms of systems memory consolidation during sleep », *Nat Neurosci*, vol. 22, nº 10, p. 1598-1610, oct. 2019, doi: 10.1038/s41593-019-0467-3.
[13] M. Mölle, T. O. Bergmann, L. Marshall, et J. Born, « Fast and slow spindles during the sleep slow oscillation: disparate coalescence and engagement in memory processing », *Sleep*, vol. 34, nº 10, p. 1411-1421, 2011, doi: 10.5665/SLEEP.1290.
[14] R. Cox, A. C. Schapiro, D. S. Manoach, et R. Stickgold, « Individual Differences in Frequency and Topography

of Slow and Fast Sleep Spindles », *Front Hum Neurosci*, vol. 11, p. 433, 2017, doi: 10.3389/fnhum.2017.00433.
[15] M. Schabus, P. Maquet, et others, « Hemodynamic cerebral correlates of sleep spindles during human non-rapid eye movement sleep », *Proceedings of the National Academy of Sciences*, vol. 104, nº 32, p. 13164-13169, 2007, doi: 10.1073/pnas.0703084104.
[16] J. Cartailler, P. Parutto, C. Touchard, F. Vallée, et D. Holcman, « Alpha rhythm collapse predicts iso-electric suppressions during anesthesia », *Communications biology*, vol. 2, nº 1, p. 1-10, 2019.
[17] C. Aldecoa *et al.*, « Update of the European Society of Anaesthesiology and Intensive Care Medicine evidence-based and consensus-based guideline on postoperative delirium in adult patients », *Eur J Anaesthesiol*, vol. 41, nº 2, p. 81-108, févr. 2024, doi: 10.1097/EJA.0000000000001876.
[18] M. Le Van Quyen et A. Bragin, « Analysis of dynamic brain oscillations: methodological advances », *Trends in neurosciences*, vol. 30, nº 7, p. 365-373, 2007.
[19] D. L. Donoho et J. M. Johnstone, « Ideal spatial adaptation by wavelet shrinkage », *Biometrika*, vol. 81, nº 3, p. 425-455, sept. 1994.
[20] I. M. Johnstone et B. W. Silverman, « Wavelet Threshold Estimators for Data with Correlated Noise », *Journal of the Royal Statistical Society: Series B (Statistical Methodology)*, vol. 59, nº 2, p. 319-351, juin 1997, doi: 10.1111/1467-9868.00071.
[21] P. Flandrin, M. Amin, S. McLaughlin, et B. Torrésani, « Time-frequency analysis and applications », *IEEE signal processing magazine*, vol. 30, nº 6, p. 19, 2013.
[22] S. Jaffard 1962-, *Wavelets: tools for science & technology*. Philadelphia, PA: Society for Industrial and Applied Mathematics, 2001.
[23] S. Jaffard, « Beyond Besov Spaces Part 1: Distributions of Wavelet Coefficients », *Journal of Fourier Analysis and Applications*, vol. 10, nº 3, p. 221-246, mai 2004, doi: 10.1007/s00041-004-0946-z.
[24] P. Ciuciu, P. Abry, C. Rabrait, et H. Wendt, « Log Wavelet Leaders Cumulant Based Multifractal Analysis of EVI fMRI Time Series: Evidence of Scaling in Ongoing and Evoked Brain Activity », *IEEE Journal of Selected Topics in Signal Processing*, vol. 2, nº 6, p. 929-943, déc. 2008, doi: 10.1109/JSTSP.2008.2006663.
[25] C. Sun et D. Holcman, « Combining transient statistical markers from the EEG signal to predict brain sensitivity to general anesthesia », *Biomedical Signal Processing and Control*, vol. 77, p. 103713, 2022.
[26] C. Sun, D. Longrois, et D. Holcman, « Spectral EEG correlations from the different phases of general anesthesia », *Frontiers in Medicine*, vol. 10, p. 1009434, 2023.
[27] M. Dora, S. Jaffard, et D. Holcman, « The WQN algorithm to adaptively correct artifacts in the EEG signal », *Applied and Computational Harmonic Analysis*, vol. 61, p. 347-356, 2022.
[28] M. Dora et D. Holcman, « Adaptive single-channel EEG artifact removal for real-time clinical monitoring », *IEEE Transactions on Neural Systems and Rehabilitation Engineering*, 2022.
[29] P. Flandrin, G. Rilling, et P. Goncalves, « Empirical mode decomposition as a filter bank », *IEEE signal processing letters*, vol. 11, nº 2, p. 112-114, 2004.
[30] N. E. Huang *et al.*, « The empirical mode decomposition and the Hilbert spectrum for nonlinear and non-stationary time series analysis », *Proceedings of the Royal Society of London. Series A: mathematical, physical and engineering sciences*, vol. 454, nº 1971, p. 903-995, 1998.
[31] W. G. Jacoby, « Loess:: a nonparametric, graphical tool for depicting relationships between variables », *Electoral studies*, vol. 19, nº 4, p. 577-613, 2000.
[32] S. M. Lundberg et S.-I. Lee, « A Unified Approach to Interpreting Model Predictions », dans *Advances in Neural Information Processing Systems*, Curran Associates, Inc., 2017. Consulté le: 18 décembre 2024. [En ligne]. Disponible à: https://papers.nips.cc/paper_files/paper/2017/hash/8a20a8621978632d76c43dfd28b67767-Abstract.html
[33] A. L. Goldberger *et al.*, « PhysioBank, PhysioToolkit, and PhysioNet: components of a new research resource for complex physiologic signals », *circulation*, vol. 101, nº 23, p. e215-e220, 2000.
[34] M. Dora, S. Jaffard, et D. Holcman, « The WQN algorithm for EEG artifact removal in the absence of scale invariance », *IEEE Transactions on Signal Processing*, 2024.
[35] R. F. Helfrich, B. A. Mander, W. J. Jagust, R. T. Knight, et M. P. Walker, « Old brains come uncoupled in sleep: slow wave-spindle synchrony, brain atrophy, and forgetting », *Neuron*, vol. 97, nº 1, p. 221-230, 2018.
[36] P. Berens, « CircStat: a MATLAB toolbox for circular statistics », *Journal of statistical software*, vol. 31, p. 1-21, 2009.
[37] M. A. Kramer et U. T. Eden, *Case studies in neural data analysis: a guide for the practicing neuroscientist*. MIT Press, 2016.
[38] J. D. Gibbons et S. Chakraborti, *Nonparametric statistical inference: revised and expanded*. CRC press, 2014.
[39] T. S. Wildes *et al.*, « Effect of electroencephalography-guided anesthetic administration on postoperative delirium among older adults undergoing major surgery: the ENGAGES randomized clinical trial », *Jama*, vol. 321, nº 5, p. 473-483, 2019.
[40] M. Wenzel, A. Leunig, S. Han, D. S. Peterka, et R. Yuste, « Prolonged anesthesia alters brain synaptic architecture », *Proc Natl Acad Sci U S A*, vol. 118, nº 7, p. e2023676118, févr. 2021, doi: 10.1073/pnas.2023676118.